\documentclass[12pt]{article}
\newcommand{\ket}[1]{|#1\rangle} 
\newcommand{\bra}[1]{\langle #1|}
\usepackage{bbm,epsf,amsmath, amsthm, amssymb}

\newtheorem{theorem}{Theorem}
\newtheorem{definition}[theorem]{Definition}
\newtheorem{lemma}[theorem]{Lemma}

\title{Symmetric measurements attaining the accessible information}
\author{Thomas Decker \\ 
{\small IAKS, Arbeitsgruppe Quantum Computing}\\
{\small Universit{\"a}t Karlsruhe (TH)}\\ 
{\small Am Fasanengarten 5, D-76\,131 Karlsruhe, Germany}\\
{\small decker@ira.uka.de}}
\date{September 19, 2005}
\begin{document}
\maketitle

\begin{abstract}
A theorem of Davies states that for symmetric quantum states there
exists a symmetric POVM maximizing the mutual information. To apply
this theorem the representation of the symmetry group has to be
irreducible. We obtain a similar yet weaker result for reducible
representations. We apply our results to the double trines ensemble
and show numerically that for this ensemble the pretty good
measurement is optimal.
\end{abstract}

%
%
\section{Introduction}
One of the basic problems of quantum information theory is the quantum
detection problem: Given an unknown element of a finite set of
possible states we want to obtain as much knowledge as possible about
this state by performing measurements. More precisely, we look for a
positive operator-valued measure (POVM) that minimizes or maximizes a
certain optimality criterion. There are different criteria for the
detection of quantum states. For example, we can consider the
detection error probability, Bayes costs~\cite{Helstrom}, or the
mutual information~\cite{Davies, Sasaki}. In this article we only
consider the mutual information of a measurement.

Compared to other criteria the mutual information leads to very hard
optimization problems even for simple state sets. This is due to the
logarithm in the definition of the mutual information whereas other
criteria as the error probability or Bayes costs are much simpler.
There is only little known about optimal measurements for the mutual
information~\cite{Davies, Sasaki,Shor,Shor2,PeresWootters}.  The
principal idea for obtaining these results is to use the convex
structure of the POVMs and the mutual information. Standard arguments
for convex functions and sets~\cite{Gruenbaum, Alfsen, Bertsimas}, e.g.,
Carath\'eodory's theorem, can be applied. Davies showed with these
arguments that we can find optimal measurements with a certain number
of POVM operators~\cite{Davies}.  Furthermore, for symmetric state
sets there exists an optimal measurement whose POVM operators
constitute a single orbit. The proof of this theorem only works for
irreducible representations of the symmetry group and the theorem
cannot be generalized directly to reducible representations as the
example of Refs.~\cite{Shor, Shor2} shows.  This means that for
certain state sets with a reducible representation of the symmetry
group it is not possible to obtain an optimal POVM which is a single
orbit.

In this article we generalize Davies' theorem to reducible
representations. The generalization states that there is an optimal
symmetric POVM where we know an upper bound for the number of
orbits. The upper bound depends on the number of irreducible
components in the representation of the symmetry group. We apply the
generalization to the double trine ensemble and show numerically that
the pretty good measurement of Refs.~\cite{PeresWootters,Wootters} is
an optimal measurement for this state set.

We proceed as follows. In the next two sections we recapitulate basic
definitions and properties of POVMs, symmetric matrices, and the
mutual information. In Sec.~\ref{Sec 4} we show how both of Davies'
theorems can be proved with a theorem that directly
follows from the theory of convex sets. This theorem leads to the
generalization to reducible representations.  In Sec.~\ref{Sec 5} we
apply the generalized theorem to two special cases of the lifted
trines.

%
%
\section{Symmetric states, POVMs, and matrices}
In this section we outline basic definitions of symmetric quantum
states and POVMs. We show that the symmetry of POVMs naturally leads
to matrices with symmetry.

\subsection{Symmetric states and POVMs}\label{Sec 2.1}
We consider a quantum system with corresponding Hilbert space
${\mathbbm C}^d$. A state of the system can be described by a density
matrix $\rho \in {\mathbbm C}^{d \times d}$, i.e., a semi-positive
matrix with ${\rm tr}(\rho)=1$. In the following we refer to a state
set\footnote{We allow multiple copies of elements in a set of
states or POVM operators, i.e., we consider multisets.}
$S=\{\rho_1, \ldots, \rho_m\}$ with corresponding prior
probabilities $p(i)$ as an {\em ensemble}. A pure state
$\rho_i=\ket{\Psi_i}\bra{\Psi_i}$ can be described by the state vector
$\ket{\Psi_i}$.  A POVM measurement is defined by a set $P=\{\Pi_1,
\ldots , \Pi_n\} \subseteq {\mathbbm C}^{d \times d}$ of non-zero
semi-positive matrices with $\sum_i \Pi_i=I_d$ where $I_d$ denotes the
identity matrix of size $d\times d$. The result of a measurement is an
index $i$ which occurs with the probability ${\rm tr}(\Pi_i \rho)$
when $\rho$ is the given state.

The symmetry of ensembles and POVMs is defined by the invariance 
of the corresponding set of matrices under the action of a group.
\begin{definition}
Let $X=\{ X_1, \ldots , X_m \} \subseteq {\mathbbm C}^{d \times d}$ be
the set of matrices corresponding to a POVM or ensemble.
Furthermore, let $G$ be a finite group with unitary
representation $\sigma:G \to {\mathbbm C}^{d \times d}$. The POVM
or ensemble is symmetric with respect to $\sigma$ 
if $X$ is invariant under the operation $X_i \mapsto 
\sigma(g) X_i \sigma(g)^\dagger$ for all $g \in
G$ and $i \in \{1, \ldots, m\}$, i.e., this operation defines a 
permutation representation on $X$. For ensembles we additionally 
assume equal prior probabilities for states of the same orbit.
\end{definition}

Following this definition, we assume that $\sigma$ is a
non-projective representation.  As discussed in Ref.~\cite{altesPapier}, a
projective representation can be transformed into a non-projective
representation by a central extension of $G$. Furthermore, we do not
assume that $G$ operates transitively on $X$. This allows 
that we can consider the symmetries that are defined by
the subgroups of $G$, too. In particular, the group $G$ can be
the trivial group.

An important construction for POVMs is the symmetrization. This
means that a POVM can be extended to a symmetric POVM as the
following lemma states~\cite{Davies}. The symmetric POVM can contain
several orbits and the matrices need not be distinct.
\begin{lemma}\label{Lemma 2}
Let $P \subseteq {\mathbbm C}^{d \times d}$ be a POVM and 
$\sigma: G\to {\mathbbm C}^{d \times d}$ a unitary 
representation of the finite group $G$. Then
$$P^G:= \left\{ \frac{1}{|G|} \sigma(g) \Pi \sigma(g)^\dagger :
g \in G, \Pi \in P\right\}$$ is a symmetric POVM.
\end{lemma}
Another construction to obtain new POVMs is the convex combination:
\begin{definition}\label{Def 3}
Let $P=\{\Pi_1, \ldots, \Pi_m\}$ and ${\tilde P}= 
\{ {\tilde \Pi}_1,\ldots, {\tilde \Pi}_n\}$ be two POVMs of a system.  
For $\lambda \in
[0,1]$ define the convex combination
$$\lambda P + (1-\lambda){\tilde P}:= \{ \lambda \Pi_1,
\ldots, \lambda \Pi_m, (1-\lambda){\tilde \Pi}_1, \ldots,
(1-\lambda){\tilde \Pi}_n\}.$$
\end{definition}
This convex combination corresponds to a random selection between two
POVMs. We do not forget which POVM we have chosen after the
measurement, i.e., we assume that the results of both POVMs are
distinct.

\subsection{Matrices with symmetry}
The matrices of a POVM are Hermitian. The $d^2$ matrices
\begin{equation}\label{Matrizen}
E_{kk} := \ket{k}\bra{k},\;
X_{kl} := \ket{k}\bra{l} + \ket{l}\bra{k}, \;{\rm and}\;
Y_{kl} := i \ket{k}\bra{l} -i \ket{l}\bra{k},
\end{equation}
$k>l$, of size $(d \times d)$ constitute an orthogonal 
basis for the real linear space of Hermitian matrices
with the trace inner product. For symmetric POVMs we construct
specific matrices in subspaces that can be described
by the theory of symmetric matrices~\cite{Egner}.
\begin{definition}\label{Def-symm}
Let $G$ be a finite group with representations $\sigma: G \to
{\mathbbm C}^{m \times m}$ and $\tau: G \to {\mathbbm C}^{n \times
n}$. The matrix $M \in {\mathbbm C}^{m \times n}$ has the symmetry
$(G,\sigma,\tau)$ if $\sigma(g) M = M \tau(g)$ for all $g \in G$. We
write $\sigma M = M \tau$.
\end{definition}
Due to Schur's lemma~\cite{Serre} a symmetric matrix has a special
structure which can be described with the intertwining
space~\cite{Pueschel} of two representations.
\begin{definition}
Let $G, \sigma$ and $\tau$ be as in Def.~\ref{Def-symm}.  The
intertwining space of $\sigma$ and $\tau$ is the linear space ${\rm
Int}(\sigma, \tau):= \{ M \in {\mathbbm C}^{m \times n} : \sigma M = M
\tau \}$.
\end{definition}
A matrix $M$ has the symmetry $(G,\sigma, \tau)$ if and only if $M \in
{\rm Int}(\sigma,\tau)$. Hence, the structure of a symmetric
matrix is determined by the structure of the intertwining
space. The latter can be easily described if we assume that
\begin{equation}\label{zerlegung}
\sigma=\bigoplus_{i=1}^z ( I_{m_i} \otimes \kappa_i) \quad {\rm and}
\quad \tau=\bigoplus_{i=1}^z ( I_{n_i} \otimes \kappa_i)
\end{equation}  are
decompositions of $\sigma$ and $\tau$ into the irreducible
representations $\kappa_i$ of $G$. These decompositions can be
obtained by conjugation of $\sigma$ and $\tau$ with appropriate
unitary matrices~\cite{Serre}.  The natural numbers $m_i$ and $n_i$ are the
multiplicities~\cite{Dornhoff} of the irreducible 
representations $\kappa_i$ in $\sigma$ and $\tau$.  The following
lemma specifies the structure of the intertwining space~\cite{Pueschel}.
\begin{lemma}\label{intertwiner}
Let $\sigma$ and $\tau$ be two representations of $G$ with the
decompositions of Eq.~(\ref{zerlegung}).  Then $${\rm Int}(\sigma,
\tau)= ({\mathbbm C}^{m_1 \times n_1} \otimes I_{{\rm deg}(\kappa_1)})
\oplus \ldots \oplus ({\mathbbm C}^{m_z \times n_z} \otimes I_{{\rm
deg}(\kappa_z)})$$ where ${\rm deg}(\kappa_i)$ denotes the degree of
$\kappa_i$.
\end{lemma}
For $m_i=0$ we insert $n_i\,{\rm deg}(\kappa_i)$ zero columns and for 
$n_i=0$ we insert $m_i\,{\rm deg}(\kappa_i)$ 
zero rows at the appropriate positions. For symmetric ensembles
and POVMs we only need a special case of this lemma. Let $X_i$ be
symmetric states or POVM operators.  Then $C:=\sum_i X_i$ is invariant
under the conjugation with $\sigma$, i.e.,
$$\sigma(g) C \sigma(g)^\dagger = C \quad \hbox{for all} \quad g \in
G.$$ This means, that $\sigma C = C \sigma$. Using
Lemma~\ref{intertwiner} we see that $C$ is a Hermitian block-diagonal
matrix with blocks that are Hermitian matrices, too.  The following
lemma determines the dimension of the intertwining space.
\begin{lemma}\label{Lemma 7}
Let $\sigma$ be as in Eq.~(\ref{zerlegung}). Then the
Hermitian matrices in ${\rm Int}(\sigma,\sigma)$ constitute a linear
space of real dimension $\sum_i m_i^2$.
\end{lemma}
Assume that $\sigma$ is irreducible. Then ${\rm Int}(\sigma,\sigma)$ is an
one-dimensional space since it contains only real scalar multiples of
the identity matrix. For a representation of the trivial group
${\rm Int}(\sigma,\sigma)$ is the full space of matrices, i.e., 
the linear space has the dimension $d^2$.

%
%
\section{Basic properties of mutual information}
Let $S$ be an ensemble and $P$ be a POVM as defined in Sec.~\ref{Sec 2.1}.
Using the conditional probability $p(j|i):={\rm tr}(\Pi_j \rho_i)$ 
we can define the joint
probability distribution $p_{ij}:=p(i)p(j| i)$.  With this
distribution we can define the mutual information as in classical
information theory~\cite{CoverThomas}.
\begin{definition}
The mutual information of the ensemble $S$ and POVM $P$ is
\begin{equation}\label{MutInf}
I(S,P):=\sum_{i=1}^m \sum_{j=1}^n H(p_{ij}) -
\sum_{i=1}^m H\left( \sum_{j=1}^n p_{ij}\right) - 
\sum_{j=1}^n H\left(\sum_{i=1}^m p_{ij}\right)
\end{equation}
with $H(u)=u\,{\rm log}_2 \, u$.
\end{definition}
The fundamental problem is to find a POVM $P$ that maximizes $I(S,P)$
for a given ensemble $S$ with prior probabilities $p(i)$.  The
information obtained by an optimal measurement is called the accessible
information~\cite{Sasaki}.

We resume some properties of the mutual information which can be
used to transform optimal measurements into a normal form.  Then the
optimization can be restricted to these POVMs. The first lemma
directly follows from classical information theory (see Th.~2.7.4 
in Ref.~\cite{CoverThomas})
and essentially states that the mutual information is a convex
function in the conditional probability $p(j|i)$ for a fixed
distribution $p(i)$.
\begin{lemma}\label{Lemma 9}
Let $P=\{\Pi_1, \ldots, \Pi_n\}$ as well as ${\tilde P}=\{{\tilde
\Pi}_1, \ldots, {\tilde \Pi}_n\}$ be POVMs and $S$ an ensemble.
Define the POVM $Q:=\{\lambda \Pi_i + (1-\lambda) {\tilde \Pi}_i : i
=1, \ldots, n\}$ for $\lambda \in [0,1]$.  Then the inequality
$$I(S,Q) \leq \lambda I(S,P) + (1-\lambda) I(S,{\tilde P})$$
holds. The equality holds if and only if
$$ p_{ij}  \sum_k {\tilde p}_{kj} = {\tilde p}_{ij}\sum_k  p_{kj}$$ for 
all $i$ and $j$ where ${\tilde p}_{ij}:= 
p(i){\rm tr}({\tilde \Pi}_j \rho_i)$. 
\end{lemma}
The equality condition holds exactly for POVMs with the property that
the probability vectors $(p_{1j}, \ldots, p_{mj})$ and $({\tilde
p}_{1j}, \ldots, {\tilde p}_{mj})$ that are induced by $\Pi_j$ and
${\tilde \Pi}_j$ for the given ensemble are equal up to a constant
factor:$$c_j (p_{1j}, \ldots, p_{mj}) = d_j({\tilde p}_{1j}, \ldots,
{\tilde p}_{mj})$$ for $c_j,d_j \geq 0$ with $c_j + d_j >0$ for each
$j$. In other words, for the given ensemble both operators are
indistinguishable up to the constant factor.

The convex combination of POVMs in Lemma~\ref{Lemma 9} differs
from Def.~\ref{Def 3}. We obtain the latter by padding $P$ from the
right and ${\tilde P}$ from the left with zero operators in such a way
that all combinations $\lambda \Pi_i + (1-\lambda){\tilde \Pi}_i$
encompass one zero operator. Then we have $\sum_k p_{kj}=0$ or $\sum_k
{\tilde p}_{kj} =0$ for all $j$, i.e., we have $c_j=0$ or $d_j=0$.
Consequently, the information obtained by the convex combination of
two POVMs is the convex combination of the corresponding informations:
\begin{lemma}\label{Lemma 10}
Let $S$ be an ensemble and let $P$ as well as ${\tilde P}$ be
POVMs. Then for all $\lambda \in [0,1]$ the equality $$I(S,\lambda P +
(1-\lambda) {\tilde P}) = \lambda I(S,P) + (1-\lambda) I(S,{\tilde
P})$$ holds.
\end{lemma}
This lemma has a simple interpretation: For measurements we randomly
choose between two devices. Then the total information we obtain is
the weighted average of the informations for each device.

The next lemma (see Lemma~2 of Ref.~\cite{Davies})
shows that POVM operators that
are equal up to normalization can be merged without changing
the mutual information of the POVM. The same
is true if we split an operator $\Pi_i$ into $\lambda \Pi_i$ and
$(1-\lambda)\Pi_i$ for $0 \leq \lambda \leq 1$. 
This theorem can be applied repeatedly and to permutations
of the operators, too.
\begin{lemma}\label{aufteilen}
Let $S$ be an ensemble. 
Then $I(S,{\tilde P})= I(S,P)$ holds for the
POVMs $P=\{\Pi_1, \Pi_2, \ldots, \Pi_n \}$ and 
${\tilde P}=\{\lambda \Pi_1, (1-\lambda) \Pi_1,\Pi_2 , \Pi_3,
\ldots, \Pi_n\}$ with $0 \leq \lambda \leq 1$. 
\end{lemma}
The optimization of POVMs can be simplified in some cases if we use a
special normalization of the POVM operators. The following definition
shows how a POVM can be rewritten in such a way that the resolution
$\sum_i \Pi_i = I_d$ of the identity is a convex
combination~\cite{Davies}.
\begin{definition}\label{Def 12}
Let $P=\{\Pi_1, \ldots, \Pi_n \}$ be a POVM with non-zero operators. 
Then write $P =
\{\lambda_1 \Pi_1^\prime, \ldots, \lambda_n \Pi_n^\prime\}$ with
$$\Pi_i^\prime := \frac{d}{{\rm tr}(\Pi_i)}\Pi_i \quad {\rm and} \quad
\lambda_i := \frac{{\rm tr}(\Pi_i)}{d}.$$ The identity is
the convex combination $\sum_i \lambda_i \Pi_i^\prime = I_d$.
\end{definition}

The last lemma of this section is a generalization of
Lemma~5 in Ref.~\cite{Davies} and can be applied to
symmetric ensembles. It states that for a given POVM the
symmetrization of this POVM has the same mutual information. Hence, we
know that for symmetric ensembles there always exists an optimal
symmetric POVM.
In the next section we prove the existence of a symmetric POVM
where we know an upper bound for the number of orbits.
\begin{lemma}\label{Lemma 14}
Let $S$ be a symmetric ensemble with symmetry group $G$ and
let $P$ be a POVM. Then $I(S,P)=I(S,P^G)$.
\end{lemma}
To apply this theorem the symmetry group need not operate transitively
on the ensemble. The probabilities $p(i)$ have to be
constant on each orbit.

%
%
\section{Optimal POVMs for symmetric ensembles}\label{Sec 4}
In the literature, the main tools for the optimization of POVMs are
Davies' theorems~\cite{Davies} and their real versions~\cite{Sasaki}.
We briefly recapitulate the proofs and generalize the theorem for
symmetric ensembles to reducible representations of the symmetry
group.

Davies' first theorem (Th.~3 of Ref.~\cite{Davies}) states that for an
ensemble of a $d$-dimensional Hilbert space there exists an optimal
POVM with $n$ rank-one operators where $d \leq n \leq d^2$.  
Davies' proof is essentially based on the following lemma
which deals with convex combinations of the identity. The
set of these combinations is convex and the lemma gives
an upper bound for the number of operators of an extreme 
point~\cite{Davies}.
We prove this lemma in the appendix with standard
arguments of linear optimization.
\begin{lemma}\label{Th 15}
Let $\sum_i \lambda_i \Pi_i^\prime = I_d$ be a convex combination with
${\rm tr}(\Pi_i^\prime)=d$. Furthermore, let all $\Pi_i^\prime$ be
elements of the affine space $I_d+K$ where $K$ is an $r$-dimensional 
linear subspace of Hermitian matrices.
Then the convex combination can be rewritten as
$$\sum_i \lambda_i \Pi_i^\prime =\sum_i \mu_i \left( \sum_j 
\nu_{ij} \Pi_j^\prime \right) \; {\rm with}\; \sum_j \nu_{ij} 
\Pi_j^\prime =I_d$$ where
$\mu_i, \nu_{ij} \geq 0$ and $\sum_j \mu_j = \sum_j \nu_{ij}=1$ for
all $i$. Furthermore, for each $i$ at most 
$r+1$ elements $\nu_{ij}$ are non-zero. 
\end{lemma}
Using this lemma we can prove the upper bound of Davies' first theorem
as follows. Assume that $\Pi_1, \ldots, \Pi_n$ is an optimal POVM.  We
can assume that it consists of rank-one operators~\cite{Davies}. Using
the normalization of Def.~\ref{Def 12} we have a convex combination
$\sum_i \lambda_i \Pi_i^\prime = I_d$.  With Lemma~\ref{Th 15} the
POVM is a convex combination of POVMs with at most $d^2$ operators
each because $K$ has dimension\footnote{The trace normalization reduces
the dimension $d^2$ of the space of Hermitian matrices by one.}
$d^2-1$. Lemmas~\ref{Lemma 10} and~\ref{aufteilen} 
show that at least one of these POVMs
is optimal, too.

We show how Davies' second theorem (Th.~4 of Ref.~\cite{Davies})
follows from Lemma~\ref{Th 15}. The former states that for a symmetric ensemble
with irreducible representation $\sigma$ there exists an optimal
POVM which is a single orbit.  Let $P$ be an optimal POVM with rank-one
operators. Following Lemma~\ref{Lemma 14} the POVM $P^G$ is optimal,
too. We consider the orbits
$$C_i:= \left\{ \frac{1}{|G|} \sigma(g) \Pi_i^\prime \sigma(g)^\dagger
: g \in G\right\}$$ of the operators of $P$.  We have the convex
combination $\sum_i \lambda_i C_i = P^G$ with $\lambda_i := {\rm
tr}(\Pi_i) / d$ as in Def.~\ref{Def 12}.  We consider the orbit sums
$$D_i:= \frac{1}{|G|} \sum_{g \in G} \sigma(g) \Pi_i^\prime
\sigma(g)^\dagger.$$ Since $P^G$ is a POVM the equation $\sum_i
\lambda_i D_i = I_d$ holds, i.e., the identity matrix is a convex
combination of the matrices $D_i$. We use the irreducibility of
$\sigma$ and obtain due to $\sigma D_i = D_i \sigma$ the equation
$D_i=I_d$. In other words, the matrices $D_i$ are elements of the
intertwining space ${\rm Int}(\sigma,\sigma)={\mathbbm C}I_d$. Since
the matrices $D_i$ have trace $d$ they are elements of the affine
space $I_d + \{0\}$ whose real dimension is $r=0$.  Following
Lemma~\ref{Th 15} there exists a convex combination of $I_d$ with a
single $D_i$. With the same arguments as for the proof of Davies'
first theorem a single orbit is sufficient for an optimal measurement.

It is clear how this proof of Davies' second theorem is modified for
reducible representations: The matrices $D_i$ are elements of the
intertwining space ${\rm Int}(\sigma,\sigma)$ which has dimension
$r:=\sum_i m_i^2$ as stated in Lemma~\ref{Lemma 7}.  The trace
normalization reduces the dimension by one. Then Lemma~\ref{Th 15}
states that we need at most $r$ orbits to construct the identity
matrix.\footnote{A consequence of the decomposition of the
POVM is that some operators are decomposed into several
copies. Lemma~\ref{aufteilen} states that this does not change the
mutual information.}  The preceding discussion shows the following
theorem.
\begin{theorem}\label{Th 16}
Let $S$ be a symmetric ensemble with $\sigma$ as defined in
Eq.~(\ref{zerlegung}). Then there exists an optimal measurement with
rank-one operators which is the union of at most $\sum_i m_i^2$
orbits.
\end{theorem}
The theorem can also be applied if we restrict the symmetry to
subgroups of the symmetry group since the action of the group must not
be transitive. However, by this reduction the bound on the number of
orbits becomes weaker since the number of different irreducible
representations decreases while the multiplicities $m_i$
increase. Therefore, for the solution of optimization problems it is
beneficial to take as much symmetry as possible. As an extreme case, 
this theorem can be applied to a representation of the trivial 
group.\footnote{Since each orbit under this symmetry comprises a
single state the prior probabilities of the states can be chosen
arbitrarily.} Then we have $m_1=d$ for the only irreducible representation
$g \mapsto (1)$ leading to the upper bound $d^2$. 
This discussion shows that Davies' first theorem can be obtained as 
special case of the generalized theorem.

Both theorems have real versions~\cite{Sasaki}.  The bound of the
first theorem can be tightened to $n \leq d(d+1)/2$ since we can
transform an optimal POVM into a POVM with real operators. Hence, the
subspace $K$ of Lemma~\ref{Th 15} does not contain linear combinations
of the elements $Y_{kl}$ of Eq.~(\ref{Matrizen}).  Additionally, the
discussion for the second theorem is also valid if we replace the
$\kappa_i$ of Eq.~(\ref{zerlegung}) with the real irreducible
representations. We obtain the upper bound $\sum_i m_i (m_i+1)/2$
where the $m_i$ are the multiplicities of the real irreducible
representations.

%
%
\section{Examples}\label{Sec 5}
We apply the real version of Th.~\ref{Th 16} to the following
ensembles in order to obtain optimal POVMs: an ensemble of slightly
lifted trines and the double trines. The theorem leads to an
optimization problem that is a special case of those in 
Refs.~\cite{Shor, Shor2}.  We identify optimal POVMs and 
discuss their properties. For
the slightly lifted trines we conclude as in Refs.~\cite{Shor, Shor2}
that a symmetric optimal POVM must at least comprise two orbits.
For the double trines we obtain an optimal POVM consisting of a single
orbit.

\subsection{Lifted trines}\label{Sec Lift}
For each $\alpha \in [0,1]$ the three vectors
$$\left(\begin{array}{c}\sqrt{\alpha}\\ \sqrt{1-\alpha}\\0\\ 
\end{array}\right), \left(\begin{array}{c} \sqrt{\alpha} \\ 
-\frac{1}{2} \sqrt{1-\alpha}\\ \frac{\sqrt{3}}{2} \sqrt{1-\alpha}
\end{array}\right), \; {\rm and} \; \left( \begin{array}{c}
\sqrt{\alpha} \\ -\frac{1}{2}\sqrt{1-\alpha}\\ -\frac{\sqrt{3}}{2}
\sqrt{1-\alpha}\end{array}\right)$$ constitute a lifted trines
ensemble.  These ensembles are interesting since for slightly lifted
trines, i.e., $\alpha$ is next to zero, it can be numerically shown
that two orbits are necessary to obtain an optimal POVM~\cite{Shor,Shor2}. 
This shows that the direct generalization of
Davies' theorem to reducible representations is not possible and that 
the bound of Th.~\ref{Th 16} can be attained. 

We follow Refs.~\cite{Shor, Shor2} and show in detail
the analysis of optimal measurements for a special case of the lifted 
trines. The symmetry group of the lifted trines 
is generated by the rotation $$R:=\frac{1}{2}
\left( \begin{array}{ccc}2&0&0\\ 0&-1 & \sqrt{3}\\ 0& -\sqrt{3} &
-1\end{array}\right)$$ about $120$ degrees. This representation of the
symmetry group contains two inequivalent real irreducible
representations.  Each irreducible representation has the multiplicity
one.  Hence, using the real version of Th.~\ref{Th 16} we need at most
two orbits $C_1=\{\Pi, R \Pi R^2, R^2 \Pi R\}$ and $C_2=\{{\tilde
\Pi}, R {\tilde \Pi} R^2, R^2 {\tilde \Pi} R \}$ with operators $\Pi$
and ${\tilde \Pi}$ of rank one to obtain an optimal POVM. With the
normalization ${\rm tr}(\Pi)= {\rm tr}({\tilde \Pi})=1$ the convex
combination $P=\lambda C_1 + (1 - \lambda) C_2$ is a POVM for
appropriate $\Pi$, ${\tilde \Pi}$, and $\lambda \in [0,1]$. We apply
Lemma~\ref{Lemma 10} and obtain\footnote{Lemma~10 can be applied to
orbits, too. However, we must replace $\sum_j p_{ij}$ by the prior
probability $p(i)$ in Eq.~(\ref{MutInf}) since $\sum_j p_{ij}=p(i)$
need not hold for a single orbit. For an orbit which is a POVM both
definitions coincide.}  the information $I(S,P)=\lambda
I(S,C_1) + (1-\lambda) I(S,C_2)$, i.e., 
the mutual information of a convex combination of orbits is the convex 
combination of the formal mutual informations $I(S,C_1)$ and $I(S,C_2)$. 
For an operator
$\Pi=\ket{\Psi(a,b)}\bra{\Psi(a,b)}$ we use the parameterization
\begin{equation}\label{Parametrisierung}
\ket{\Psi(a,b)}=\left( \begin{array}{c}
{\rm cos}(a)\\
{\rm sin}(a){\rm cos}(b)\\
{\rm sin}(a){\rm sin}(b)\end{array}\right)
\end{equation}
leading to the orbit sum $$\sum_{i=0}^2 R^i \Pi R^{-i}
=\left( \begin{array}{ccc}3 \, {\rm cos}^2(a) &0&0\\
0&\frac{3}{2}-\frac{3}{2} \, {\rm cos}^2(a)&0 \\ 
0&0& \frac{3}{2}-\frac{3}{2}\, {\rm cos}^2(a)
\end{array}\right).$$
For two orbits $C_1$ and $C_2$ with parameters
$(a,b)$ and $(c,d)$
the convex combination  $\lambda C_1 + (1-\lambda)C_2$ is a POVM,
i.e., the sum of all operators equals $I_3$, 
if and only if
\begin{equation}\label{konvex}
\lambda \, {\rm cos}^2(a) + (1-\lambda) \, {\rm cos}^2(c) = 
\frac{1}{3}.
\end{equation}
If we assume ${\rm cos}^2(a) \leq {\rm cos}^2(c)$ this means that $1/3
\in [ {\rm cos}^2(a), {\rm cos}^2(c) ]$, i.e., ${\rm cos}^2(a) \in
[0,1/3]$ and ${\rm cos}^2(c) \in [1/3,1]$ are all possible values.  In
the following we only consider the mutual information $I(a,b)$ of the
orbit with $a={\rm arccos}\sqrt{x}$, $x \in [0,1]$, and 
$b \in [0,2\pi/3]$. This is sufficient since
for a given value ${\rm cos}(a)$ with 
${\rm cos}^2(a)=x$ we have the four possible values
${\rm cos}(a)=\pm \sqrt{x}$ and ${\rm sin}(a)=\pm \sqrt{1 - x}$
in the vector $\ket{\Psi(a,b)}$ of Eq.~(\ref{Parametrisierung}). We
denote these combinations of signs 
by $++$, $+-$, $-+$, and $--$. The case $--$
leads to the same information as $++$ since the corresponding vectors
differ only by a 
global phase. With the same argument the cases $-+$ and $+-$ lead to
the same mutual information. For $+-$ the vector has a minus sign in the
last two components. Hence, we have the same information as for 
$++$ where we replace $b$ by $b+\pi$.
This discussion shows that the optimization with
$({\rm arccos}\sqrt{x},b)$ for $x \in [0,1]$ and $b \in [0,2\pi/3]$ 
takes all possible values into account.\footnote{The values of $b$ can be 
restricted due to the symmetry.}

\begin{figure}
\centerline{\epsffile{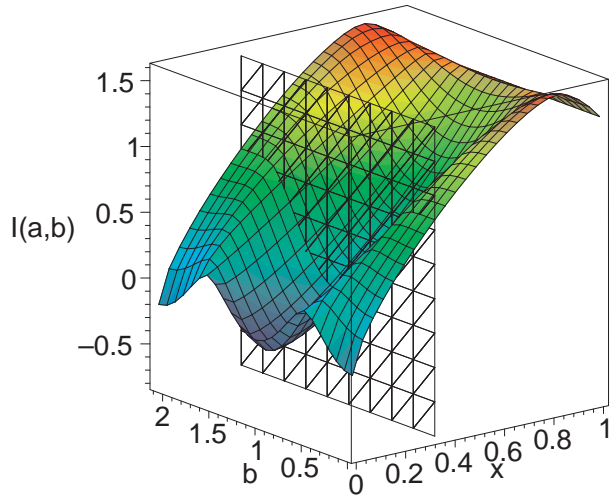}} 
\caption{The mutual information $I({\rm arccos}\sqrt{x},b)$ 
   for the lifted trines with $\alpha=1/20$. The information of an orbit
   can be negative. However, as Fig.~\ref{lifted2} shows 
   the convex combination of the information
   of two points on different sides of the plane $x=1/3$ 
   leads to a non-negative information.}
\label{lifted1}
\end{figure}

\begin{figure}
\centerline{\epsffile{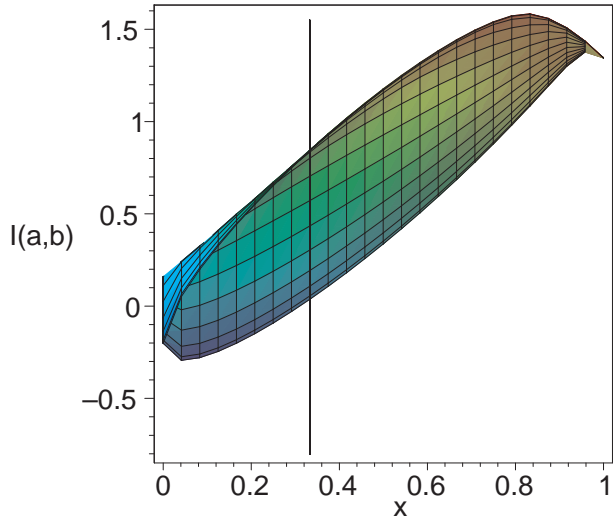}} 
\caption{Fig.~\ref{lifted1} viewed along the $b$-axis shows the
maximum of $I({\rm arccos}\sqrt{x},b)$ 
for each $x$ whereas $b$ remains a variable.
This maximum is slightly convex in the interval $[0,0.3831]$ of the
$x$-axis.
}
\label{lifted2}
\end{figure}

We restrict our attention to $\alpha=1/20$, i.e., to an example of
slightly lifted trines. In Figs.~\ref{lifted1} and \ref{lifted2} 
the information $I({\rm arccos}\sqrt{x},b)$ of an orbit with
parameters $({\rm arccos}\sqrt{x},b)$ is shown.
Due to the symmetry each probability $p_{ij}$ is equal
to the probability $p_{1k}$ for a certain $k$. 
Hence, the information of the orbit is 
$$I({\rm arccos}\sqrt{x},b)=
3(H(p_{11})+H(p_{12})+H(p_{13})-H ( p_{11}+p_{12}+p_{13})) +{\rm log}_23.$$
The condition of Eq.~(\ref{konvex}) means that
\begin{figure}
\centerline{\epsffile{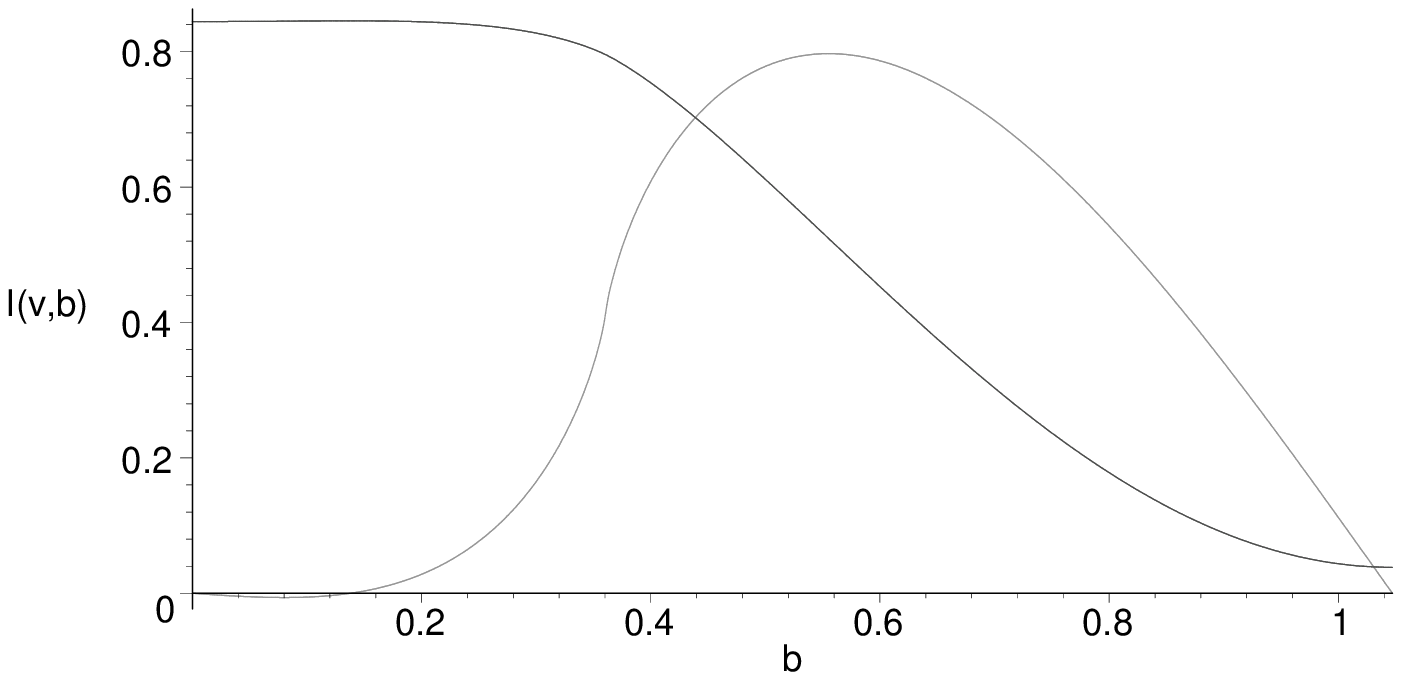}} 
\caption{The function $I(\nu,b)$ has a global maximum at $b\approx
  0.1377$. It seems to be complicated to give a simple analytic
  definition of this value. The partially negative function is 
  $-(d/d  b)I(\nu,b)/2$.  The remaining values can be obtained by virtue of the
  symmetry of the function.}
\label{lifted3}
\end{figure} 
the points $(a, b)$ and $(c, d)$ we choose for our orbits lie on
different sides of the plane $x=1/3$.  
From the figures it follows that a
single orbit cannot be optimal since a POVM with a single orbit
corresponds to a point on this plane. More
precisely, the optimal information we can obtain for points on this
plane is $0.8456$ bit as shown in Fig.~\ref{lifted3}. 
This can be obtained for the POVM with
$(\nu, b)$ where $\nu:={\rm arccos}\sqrt{1/3}$
and $b \approx 0.1377$. 
The slight convexity in Fig.~\ref{lifted2} 
as well as Refs.~\cite{Shor, Shor2}
suggest that we can obtain more
information with two points: a point with $x=0$ and a point with
$b=0$ on the other side of the plane. Numerical computations
show that an optimal point for $x=0$ is
$(\pi/2,\pi/2)$ with the information $0.15996$ bit. The other 
optimal point can be
chosen to be $(\arccos\sqrt{0.3831},0)$ with $0.9499$ bit. The
convex combination of both informations is $0.8472$ bit. This is more
than the information of the optimal single orbit. 

In the following we show that the accessible information cannot
be obtained with a POVM which is a single orbit
even if we consider operators of higher rank. 
Consequently, the characterization~\cite{DAriano,Chiribella}
of the extreme points of the 
convex set of POVMs consisting of a single orbit
cannot be applied. Assume that $P:=\{R^i \Pi R^{-i}
: i=0,1,2\}$ is an optimal POVM with initial operator
$\Pi=\sum_{i=1}^3 s_i \ket{\Psi_i}\bra{\Psi_i}$ where
$\langle \Psi_i | \Psi_i \rangle = 1$, $\sum_i s_i = 1$, and
$s_i \geq 0$. Then Lemma~2 of Ref.~\cite{Davies} 
and Lemma~\ref{Lemma 10} state that
\begin{equation}\label{ugl}
I(S,P) \leq \sum_i s_i I(S,P_i) = I(S,\sum_i s_i P_i )
\end{equation} 
with $P_i:=\{R^j \ket{\Psi_i}\bra{\Psi_i}R^{-j}: j=0,1,2\}$.  The POVM
$\sum_i s_i P_i$ consists of three orbits.  Using Th.~\ref{Th 16} we
construct an optimal POVM with two of these three orbits. Without loss
of generality we assume that the orbits correspond to 
$\ket{\Psi_1}\bra{\Psi_1}$ and $\ket{\Psi_2}\bra{\Psi_2}$. 
The two corresponding points $(a,b)$ must be the optimal 
points\footnote{The point $(\pi/2,\pi/6)$ leads to the same results
as $(\pi/2,\pi/2)$.} given above.
The probability vectors
$(0.2375,0,0.2375)$ for $\ket{\Psi_1}\bra{\Psi_1}$ and
$(0.2724,0.0199,0.0199)$ for $\ket{\Psi_2}\bra{\Psi_2}$ are not equal
up to a constant factor. Therefore, following Lemma~\ref{Lemma 9}
inequality (\ref{ugl}) is strict, i.e., the single orbit cannot be an
optimal POVM.

\subsection{Double trines}
\begin{figure}
  \centerline{\epsffile{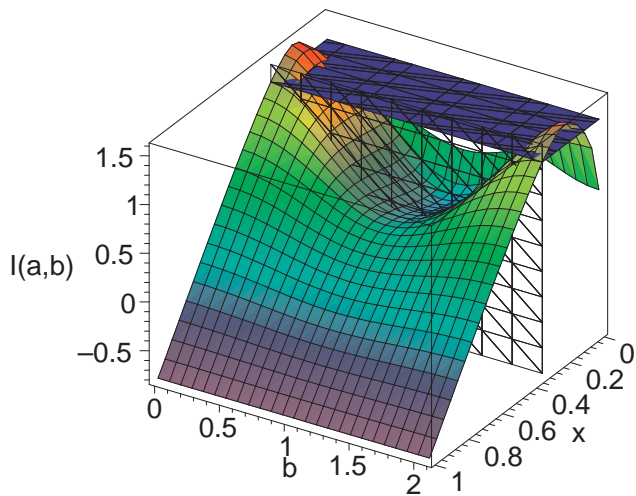}} 
  \caption{The mutual information $I({\rm arccos}\sqrt{x},b)$  
   for the double trines ensemble. The horizontal 
   plane corresponds to the information $1.369$ bit which is the information
   of the pretty good measurement.}
  \label{double1}
\end{figure}

\begin{figure}
  \centerline{\epsffile{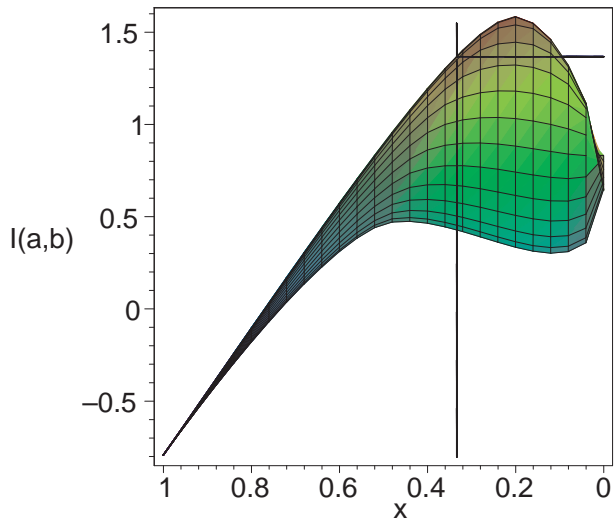}} 
  \caption{Fig.~\ref{double1} viewed along the $b$-axis shows the
   maximum of $I({\rm arccos}\sqrt{x},b)$ for each $x$ 
   whereas $b$ remains a variable.}
  \label{double2}
\end{figure}

The double trines~\cite{PeresWootters,Wootters} are defined by the three 
state vectors $$
\left(\begin{array}{c}1\\0 \end{array}\right) \otimes
\left(\begin{array}{c}1\\0 \end{array}\right) =
\left(\begin{array}{cc} 1 \cr 0 \cr 0 \cr 0
\end{array} \right), \; 
\frac{1}{2}\left(\begin{array}{c}-1 \\ -\sqrt{3} \end{array}\right) \otimes
\frac{1}{2}\left(\begin{array}{c}-1 \\ -\sqrt{3} \end{array}\right) =
\frac{1}{4}\left(\begin{array}{cc}
1\cr \sqrt{3} \cr \sqrt{3} \cr 3\end{array}\right)$$ 
$$\; {\rm and} \; \;
\frac{1}{2}\left(\begin{array}{c}-1 \\ \sqrt{3} \end{array}\right) \otimes
\frac{1}{2}\left(\begin{array}{c}-1 \\ \sqrt{3} \end{array}\right) =
\frac{1}{4}\left(\begin{array}{cc} 1\cr -\sqrt{3} \cr
-\sqrt{3} \cr 3\end{array}\right)$$ of two qubits\footnote{Compared to
the symmetry of the lifted trines in Sec.~\ref{Sec Lift}
we have the additional symmetry
operation that interchanges the qubits. Even with this operation the
representation of the symmetry group is reducible. We do not consider
this symmetry operation in the following since the decomposition of
the representation does not become simpler.}. 
We apply the unitary basis transform
$$\frac{1}{\sqrt{2}} \left(\begin{array}{cccc}
1&0&0&1\\
1&0&0&-1\\
0&1&1&0\\
0&1&-1&0\end{array}\right)$$ 
and obtain the state vectors
$$
\frac{1}{\sqrt{2}}\left(\begin{array}{c}1\\1\\0\\0\end{array}\right),\;
\frac{1}{\sqrt{2}}\left(\begin{array}{c}1\\-1/2\\ \sqrt{3}/2 \\ 0
\end{array}\right),\; {\rm and} \; 
\frac{1}{\sqrt{2}}\left(\begin{array}{c}1\\-1/2\\ -\sqrt{3}/2 \\
    0\end{array}\right).$$ We omit the last component\footnote{An
      optimal POVM operating on the four dimensions can be projected
      to a POVM on the three dimensions.  This projection does not
      change the mutual information.}  and obtain the
    lifted trines with $\alpha=1/2$. In contrast to the previous
    section these trines are strongly lifted. As mentioned in
    Refs.~\cite{Shor,Shor2} this leads to different properties of
    optimal POVMs.
We replace $\alpha=1/20$ by $\alpha=1/2$ in the computations of 
Sec.~\ref{Sec Lift} and obtain Figs.~\ref{double1} 
and~\ref{double2}
where the information $I({\rm arccos}\sqrt{x},b)$ is shown. 
An optimal POVM can be obtained with a 
convex combination of at most two orbits. It follows 
from Fig.~\ref{double2} that a single orbit with $x=1/3$ is optimal
since the convex combination of the information of two points 
on different sides of the plane $x=1/3$ is strictly below the
maximum of the information on this plane.
Computations show that $I(\nu,b)$ with $\nu={\rm arccos}\sqrt{1/3}$ 
has an extreme point at $b=0$. Hence, an optimal point on this plane is 
$(a,b)=(\nu, 0)$ leading to the information
$$I(\nu,0)= \frac{2\sqrt{2} \gamma - 9\, {\rm ln}(2)}{6\,{\rm ln}(2)}
\approx 1.369$$ with $\gamma= {\rm ln} \left(2(3+2\sqrt{2})^2 \right)$.
This  is equal
to the information obtained in
Refs.~\cite{PeresWootters,Wootters} for the pretty good
measurement~\cite{Hausladen}.  Furthermore, the Hessian 
$$\left(\begin{array}{cc}
\frac{81-27\sqrt{2}\gamma}{16 \, {\rm ln}(2)}& 0\\
0 & \frac{6-(2+\sqrt{2})\gamma}{3 \,{\rm ln}(2)}
\end{array}\right)\approx \left(\begin{array}{cc}
-7.221 & 0 \\ 0 & -4.041 \end{array}\right)$$
of $I({\rm arccos}\sqrt{x},b)$
is negative definite at the point $(x,b)=(1/3,0)$, i.e., 
the information is concave in this region.
These arguments and the global properties of 
$I({\arccos}\sqrt{x},b)$ which can be seen in Figs.~\ref{double1} 
and~\ref{double2} suggest that the POVM 
corresponding to the point $(\nu,0)$ is optimal.

%
%
\section{Conclusions}
We have generalized Davies' theorem for symmetric ensembles to
reducible representations of the symmetry group. There always exists
an optimal POVM consisting of at most $\sum_i m_i^2$ orbits where the
$m_i$ are the multiplicities of the irreducible components in the
representation of the symmetry group.

%
%
\section*{Acknowledgments}
The author acknowledges helpful discussions with D.~Janzing. 
This work was supported by Landesstiftung
Baden-W\"urttemberg gGmbH (AZ 1.1322.01).

%
%
\section*{Appendix}
In this appendix we prove Lemma~\ref{Th 15} with standard arguments of linear
optimization. Consider a POVM on a $d$-dimensional system with
operators $\Pi_1, \ldots, \Pi_n$ where we write $\Pi_i = \lambda_i
\Pi_i^\prime$ as in Def.~\ref{Def 12}. We have the convex combination $\sum_i
\lambda_i \Pi_i^\prime = I_d$.  With respect to the basis of
Eq.~(\ref{Matrizen}) this convex combination can be written as
equation
\begin{equation}\label{Gleichungssystem}
D \lambda = \left(\begin{array}{c}1\\0 \end{array}\right) \; {\rm
where} \; D:=\left( \begin{array}{ccc}1 &\ldots &1 \\ &E&\\&X& \\ &Y&
\end{array}\right).
\end{equation} 
Here we write $\lambda=(\lambda_1, \ldots, \lambda_n)^T$. The $1$
on the right side of the equation is the all-one vector of
length $d+1$ and the $0$ is the all-zero vector of length $d^2-d$.
The matrices $E \in {\mathbbm R}^{d \times n}$ and $X, Y
\in {\mathbbm R}^{ ((d^2-d)/2) \times n}$ contain the
coefficients of the real linear combinations
$$\Pi_i=\sum_{k=0}^{d-1} \Pi_i^{kk} E_{kk} +
\sum_{k > l} \Re(\Pi_i^{kl}) X_{kl} + \sum_{k > l} \Im(\Pi_i^{kl})
Y_{kl}$$ of $\Pi_i=( \Pi_i^{kl})_{kl}$ 
where $\Re$ and $\Im$ denote the real and imaginary part of 
a complex number. More precisely, the $i$-th column of $E$, $X$, and 
$Y$ contains the coefficients $\Pi_i^{kk}$, $\Re(\Pi_i^{kl})$, and
$\Im(\Pi_i^{kl})$, respectively.

We discuss some elementary properties of the solutions 
${\cal L}_E$ of $E\lambda=b$ with 
$\lambda_i \geq 0$ and the vector $b$ consisting of $d$ ones. 
The matrix $E$ contains 
only non-zero columns and non-negative entries.
\begin{lemma}\label{Lemma 17}
The set ${\cal L}_E$ is convex and compact.
\end{lemma}
\begin{proof}
The set ${\cal L}_E$ is closed and convex. Assume that it is
unbounded. Then following Th.~2.5.1 of Ref.~\cite{Gruenbaum} it
contains a ray, i.e., the set $\{
p+ \mu q : \mu \geq 0\}$ with $q \not = 0$. Choose $k$ with
$q_k\not = 0$. Since we have $\lambda_i \geq 0$ the vector $q$ cannot
contain negative entries.  For all $\mu \geq 0$ we have $E(p+\mu
q)=Ep + \mu Eq = b$.  Hence, we have $Ep=b$ and $Eq=0$. The
inequality  $Eq \geq E_k q_k$ holds 
where $E_k$ denotes the $k$-th column of $E$ and $\geq$ the
element-wise relation. Since $E$ does not contain a zero column we
have $Eq\geq E_k q_k>0$ in contradiction to $Eq=0$. Hence, the 
set ${\cal L}_E$ cannot contain a ray.
\end{proof}
This lemma can also be applied to ${\cal L}_D$ since the additional
equations restrict the set of solutions even more. Hence, ${\cal L}_D$
is convex and compact. 
\begin{lemma}\label{Lemma 18}
A compact convex subset of ${\mathbbm R}^n$ is the convex
hull of its extreme points.
\end{lemma}
\begin{proof}
See Th.~2.4.5 of Ref.~\cite{Gruenbaum}.
\end{proof}
\begin{lemma}\label{Lemma 19}
The set ${\cal L}_D$ has only a finite number of extreme points.
For an extreme point $\lambda=(\lambda_1, \ldots , \lambda_n)^T$ of 
${\cal L}_D$ we have at most ${\rm rank}(D)$ non-zero elements $\lambda_i$.
\end{lemma}
\begin{proof}
Following Th.~2.3 in Ref.~\cite{Bertsimas}
an extreme point of ${\cal L}_D$ corresponds to a feasible basic 
solution of $D\lambda=c$ where $c$ is the vector of 
Eq.~(\ref{Gleichungssystem}) consisting of ones and zeros.
Since the equation 
$\sum_i \lambda_i \Pi_i^\prime=I_d$ shows that 
${\cal L}_D$ is non-empty we can remove linear dependent rows
of $D$ without changing the set of solutions (see Th.~2.5 of
Ref.~\cite{Bertsimas}). Then Th.~2.4 of Ref.~\cite{Bertsimas} 
states that a basic solution $\lambda$ has at most ${\rm rank}(D)$ 
non-zero entries. 
The number of extreme points is finite
due to Corollary~2.1 of Ref.~\cite{Bertsimas}.
\end{proof}

In the next lemma we show that ${\rm rank}(D)$ is bounded by
the dimension of the space that contains all operators $\Pi_i^\prime$.
\begin{lemma}\label{Lemma 20}
Let $\Pi_1^\prime, \ldots, \Pi_n^\prime \in {\mathbbm C}^{d
\times d}$ with ${\rm tr}(\Pi_i^\prime)=d$ be elements of the affine
space $I_d+K$ where $K$ is a $r$-dimensional 
linear space of Hermitian matrices. Then the matrix $D$ defined in
Eq.~(\ref{Gleichungssystem}) has at most rank $r+1$.
\end{lemma}
\begin{proof}
The matrix $D$ without the first row has at most rank $r+1$ since an
affine space of dimension $r$ is contained in a linear space of
dimension $r+1$. The first row does not increase the rank since it is
linear dependent to the rows of $E$. This is due to the normalization
${\rm tr}(\Pi_i^\prime)=d$ which means that the sum of each column of $E$ is
$d$.
\end{proof}

With the lemmas of this appendix we prove Lemma~\ref{Th 15}.
\begin{proof}[Proof of Lemma~\ref{Th 15}]
As in Eq.~(\ref{Gleichungssystem}) we write $\sum_i \lambda_i
\Pi_i^\prime = I_d$ as $D\lambda=c$ with the vector $c$ of
Eq.~(\ref{Gleichungssystem})
consisting of ones and zeros. Following Lemma~\ref{Lemma 20}
the matrix $D$ has at most rank $r+1$. Then Lemma~\ref{Lemma 19} states
that an extreme point $\lambda$ of ${\cal L}_D$
has at most $r+1$ non-zero
elements. With Lemma~\ref{Lemma 17} we know that the solutions of
$D\lambda=c$ constitute a convex and compact set which is the convex
combination of its extreme points as stated in 
Lemma~\ref{Lemma 18}.
\end{proof}

%
%


\begin{thebibliography}{10}

\bibitem{Helstrom}
C.W.~Helstrom:
\newblock Quantum Detection and Estimation Theory.
\newblock Academic Press, 1976.

\bibitem{Davies}
E.B.~Davies:
\newblock Information and quantum measurement.
\newblock IEEE Inf. Theory, IT-24, 596 (1978).

\bibitem{Sasaki}
M.~Sasaki, S.M.~Barnett, R.~Jozsa, M.~Osaki, O.~Hirota:
\newblock Accessible information and optimal strategies for
real symmetrical quantum sources.
\newblock Phys. Rev. A, Vol.~59, No.~5, pp.~3325-3335, 1999.

\bibitem{Shor}
P.W.~Shor: 
\newblock On the Number of Elements Needed in a POVM Attaining the 
Accessible Information.
\newblock Quantum, Communication, Measurement and Computing 3, 
Edited by O. Hirota and P. Tombesi, Kluwer Academic, 2001.
See also quant-ph/0009077.

\bibitem{Shor2}
P.W.~Shor:
\newblock The Adaptive Classical Capacity of a Quantum Channel.
\newblock IBM Journal of Research and Development,  	
Vol.~48, No.~1, pp.~115-138, 2004.

\bibitem{PeresWootters}
A.~Peres, W.K.~Wootters:
\newblock Optimal Detection of Quantum Information.
\newblock Phys. Rev. Lett., Vol.~66, No.~9, pp.~1119-1122, 1991.

\bibitem{Gruenbaum}
B.~Gr\"unbaum:
\newblock Convex polytopes.
\newblock Wiley, 1967.

\bibitem{Alfsen}
E.M.~Alfsen:
\newblock Compact Convex Sets and Boundary Integrals.
\newblock Springer, 1971.

\bibitem{Bertsimas}
D.~Bertsimas, J.N.~Tsitsiklis:
\newblock Introduction to Linear Optimization.
\newblock Athena Scientific, 1997.

\bibitem{Wootters}
W.K.~Wootters:
\newblock Distinguishing unentangled states with an unentangled 
measurement.
\newblock quant-ph/0506149.

\bibitem{altesPapier}
T.~Decker, D.~Janzing, M.~R\"otteler:
\newblock Implementation of group-covariant positive operator
valued measures by orthogonal measurements.
\newblock J.~Math. Phys.~46, 012104~(2005).

\bibitem{Egner}
S.~Egner, M.~P\"uschel:
\newblock Symmetry-Based Matrix Factorization.
\newblock J.~Sym. Comp., Vol.~37, No.~2, pp.~157-186, 2004.

\bibitem{Serre}
J.-P.~Serre:
\newblock Linear Representations of Finite Groups.
\newblock Springer, 1977.

\bibitem{Dornhoff}
L.~Dornhoff:
\newblock Group Representation Theory, Part A.
\newblock Dekker, 1971.

\bibitem{Pueschel}
M.~P\"uschel:
\newblock Decomposing Monomial Representations of Solvable Groups.
\newblock J.~Sym. Comp., Vol.~34, No.~6, pp.~561-596, 2002.

\bibitem{CoverThomas}
T.M.~Cover, J.A.~Thomas:
\newblock Elements of information theory.
\newblock Wiley, 1991.

\bibitem{DAriano}
G.M.~D'Ariano:
\newblock Extremal covariant quantum operations and positive
operator valued measures.
\newblock J.~Math. Phys.~45, pp.~3620-3635 (2004).

\bibitem{Chiribella}
G.~Chiribella, G.M.~D'Ariano:
\newblock Extremal covariant positive operator valued measures.
\newblock J.~Math. Phys.~45, 4435 (2004).

\bibitem{Hausladen}
P.~Hausladen, W.K.~Wootters:
\newblock A `pretty good' measurement for distinguishing quantum states.
\newblock J. Mod. Opt., Vol.~41, No.~12, pp.~2385-2390, 1994.

\end{thebibliography}
\end{document}